\documentstyle[twocolumn,epsf]{jpsj}

\newcommand{\be}{\begin{equation}}
\newcommand{\ee}{\end{equation}}
\newcommand{\bea}{\begin{eqnarray}}
\newcommand{\eea}{\end{eqnarray}}
\newcommand{\lsim}{\raise.35ex\hbox{$<$}\kern-0.75em\lower.5ex\hbox{$\sim$}}
\newcommand{\gsim}{\raise.35ex\hbox{$>$}\kern-0.75em\lower.5ex\hbox{$\sim$}}
\def\runtitle{Competition between spin exchange and correlated hopping}
\def\runauthor{Y. Saiga, M. Imada}

\title
{
Competition between spin exchange and correlated hopping
}

\author
{ 
Y. Saiga$^{\rm a}$\footnote{E-mail address: saiga@stat.phys.titech.ac.jp},
M. Imada$^{\rm b}$
}

\inst
{
$^{\rm a}$Department of Physics, Tokyo Institute of Technology, 
2-12-1 Oh-okayama, Meguro-ku, Tokyo 152-8551 \\
$^{\rm b}$Institute for Solid State Physics, University of Tokyo, 
Kashiwa-no-ha 5-1-5, Kashiwa 277-8581 \\
}

\recdate
{
\hspace{2cm}
}

\abst
{
The ground-state phase diagram is numerically studied for an electronic
model consisting of the spin exchange term ($J$) and the correlated 
hopping term ($t_3$: the three-site term).
This model has no single-particle hopping and
the ratio of the two terms is controlled by a parameter 
$\alpha \equiv 4 t_3 / J$.
The case of $\alpha=1$ corresponds to complete suppression
of single-particle hopping in the strong-coupling limit 
of the Hubbard model.
In one dimension, phase separation takes place below a critical value
$\alpha_{\rm c} = 0.36$-$0.63$ which depends on
the electron density.
Spin gap opens in the whole region except the phase-separated one.
For $\alpha \gsim 1.2$ and low hole densities, charge-density-wave correlations
are the most dominant, whereas singlet-pairing
correlations are the most
dominant in the remaining region.
The possibility of superconductivity in the two-dimensional case is
also discussed, based on equal-time pairing correlations.
}

\kword
{
A. Superconductors; D. Phase transitions; D. Superconductivity
}

\begin{document}
\sloppy
\maketitle


The discovery of high-$T_{\rm c}$ cuprates~\cite{BM}
has driven us the question whether the new microscopic theory for pairing mechanism
is required.
In the cuprates strong Coulomb repulsion
may be essential,
and various pairing mechanisms have been proposed under strong correlation.
One of the common features in 
experiment is the extremely flat band
around the momenta $(0,\pi)$ and $(\pi,0)$ in units of the reciprocal lattice parameter,
which has been observed
in angle-resolved photoemission spectroscopy (ARPES)~\cite{Dessau}.
This feature is attributed to a strongly correlated effect.
The band structure similar to the ARPES data has been
numerically obtained for the Hubbard model with strong on-site Coulomb
repulsion~\cite{BSW} and
the {\it t-J} model~\cite{DNB} in two dimensions.

If the flat band plays a crucial role 
in increasing the critical temperature,
proper control of the single-particle process
should lead to more enhancement of superconducting
correlations~\cite{TI,IK,KI}.
Numerical results for one-dimensional (1D) electron systems show
that flattening of the band near the Fermi points actually extends 
the parameter region where pairing correlations are the most dominant~\cite{KI}.
If single-particle hopping is completely absent as inessentials,
other hopping mechanism is necessary
for superconductivity.
One of the candidates is the correlated hopping, 
which is often called the three-site term and is naturally derived
in the strong-coupling expansion of the Hubbard model~\cite{HL,Hirsch}.
However, in one dimension, the correlated hopping
is known to be suppress pairing correlations in the low-hole-density
region~\cite{Ammon,Batista}.
This is because the correlated hopping
leads to a repulsion between holes~\cite{Ammon}.
In contrast, spin exchange interaction favors the attractive configuration
of electrons.
Therefore the correlated hopping and the spin exchange should compete
in an electron-density region,
and the ground-state phase diagram is nontrivial.

In this paper, 
we shall clarify the ground-state properties due to the competition
between spin exchange and correlated hopping
in one dimension.
We also discuss realization of superconductivity under complete suppression
of single-particle hopping in the 2D case.
To this end, exact diagonalization is employed for finite systems.

In the Hilbert space without double occupancy of electrons,
the strong-coupling expansion of the Hubbard model
yields the following Hamiltonian
up to the order $t^2 / U$ ($U$ being the on-site Coulomb 
repulsion)~\cite{HL,Hirsch}:
\bea
  {\cal H} &=& {\cal H}_t + {\cal H}_J + {\cal H}_{t3},\label{Hamiltonian} \\
  {\cal H}_t &=& - t \sum_{\langle ij \rangle \sigma} 
\left( \tilde{c}_{i \sigma}^{\dagger} 
\tilde{c}_{j \sigma} + {\rm H.c.} \right),\label{Hamiltoniant} \\
  {\cal H}_J &=& J \sum_{\langle ij \rangle} 
\left( \mbox{\boldmath $S$}_i \cdot \mbox{\boldmath $S$}_j 
- \frac{1}{4} n_i n_j \right),\label{HamiltonianJ} \\
  {\cal H}_{t3} &=& - t_3 \sum_{\langle ij \ell \rangle \sigma}
\left( \tilde{c}_{i \sigma}^{\dagger} n_{j,-\sigma} \tilde{c}_{\ell \sigma} 
\right. \nonumber \\
  & & \left. - \tilde{c}_{i \sigma}^{\dagger} \tilde{c}_{j,-\sigma}^{\dagger}
\tilde{c}_{j \sigma} \tilde{c}_{\ell,-\sigma} + {\rm H.c.} \right),\label{Hamiltoniant3}
\eea
where $\langle ij \rangle$ and $\langle ij \ell \rangle$ are nearest neighbors,
$J = 4 t^2/U$ and $t_3 = \alpha J/4$ with $\alpha=1$.
The constrained fermion operator $\tilde{c}_{i \sigma}$ is given by
$\tilde{c}_{i \sigma} = c_{i \sigma} (1 - n_{i,- \sigma})$.
Here we consider the value of $\alpha$ as a continuous parameter.
In one dimension, the ground-state phase diagram in the $J/t$-$n$ plane
($n$ being the electron density)
has been investigated for $\alpha = 0$~\cite{OLSA} and several values of 
$\alpha > 0$~\cite{Ammon}.
We refer to a model with ${\cal H}_t$ deleted, namely
${\cal H}_{Jt3} \equiv {\cal H}_J + {\cal H}_{t3}$,
as the $J$-$t_3$ model.
We will complete a phase diagram
in the $\alpha$-$n$ plane for the $J$-$t_3$ model.
Originally, both ${\cal H}_J$ and ${\cal H}_{t3}$ decay more rapidly
than ${\cal H}_t$ when the hopping integral $t$ approaches zero.
In considering possible mechanisms for superconductivity, however,
it would be meaningful to 
deal with each term of eq.\ (\ref{Hamiltonian}) individually
away from the Hubbard model.

It should be noted that in the 1D $J$-$t_3$ model the set of all eigenvalues under
the antiperiodic boundary condition is equivalent to that under the
periodic one.

We first discuss the phase diagram in the 1D case.
In order to estimate the boundary of phase separation, 
we calculate the compressibility given by
\be
  \kappa = \frac{N}{N_{\rm e}^2} \cdot \frac{4}
  {E_0(N; N_{\rm e}+2) + E_0(N; N_{\rm e}-2) - 2 E_0(N; N_{\rm e})},
\ee
where $E_0(N; N_{\rm e})$ denotes the ground-state energy with $N_{\rm e}$ electrons
in $N$ sites.
In Fig.\ 1 we show the values of $\alpha_{\rm c}$ at which $1/\kappa$
crosses zero for $N=10$-$16$.
The size dependence is  small and this implies a reliable estimate
of the thermodynamic results.
The phase separation occurs for $\alpha < \alpha_{\rm c}$.

In the low-electron-density limit, one can estimate the phase boundary
by solving two-electron problem~\cite{EKL}.
In the $J$-$t_3$ model, the phase separation disappears at $n \sim 0$
for $\alpha > \alpha_{\rm c} = 2 \ln 2 - 1 \simeq 0.386$~\cite{Ammon}.
This estimate would be exact if there are no bound states with more than
two electrons.
In the high-electron-density limit, on the other hand,
one may estimate the boundary from balance between the gain in kinetic energy
and the loss of exchange energy due to insertion of two holes
in a Heisenberg chain~\cite{OLSA}.
Following ref.\ ~\cite{OLSA}, we calculate the ground-state energy with 
$N-2$ electrons in $N$ sites
relative to the Heisenberg energy in $N-2$ sites.
This energy becomes negative for $\alpha > 0.497, 0.488, 0.483$ and $0.480$
for $N=10, 12, 14$ and $16$, respectively.
We obtain $\alpha_{\rm c} \sim 0.47$ from extrapolation to the thermodynamic limit.
This value is smaller than $\alpha_{\rm c} \sim 0.63$ obtained
from the compressibility at the highest electron density ($n=0.875$)
in our calculation.
Anyway the critical values at $n \sim 0$ and $n \sim 1$ are both smaller 
than $\alpha = 1$,
which implies that the $\alpha = 1$ line is not interrupted by phase separation.
\begin{figure}
\vspace*{0.5cm}
\epsfxsize=7cm
\centerline{\epsfbox{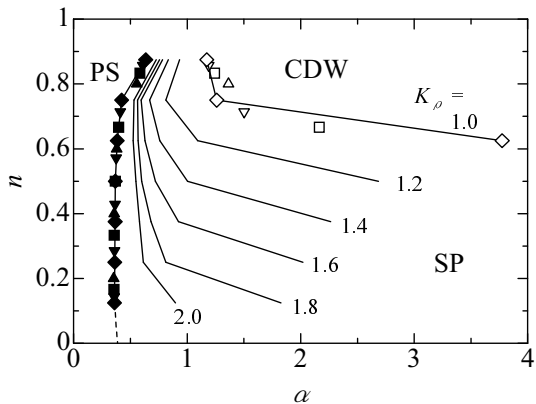}}
\caption{Phase diagram for the ground state of the 1D $J$-$t_3$ model.
PS: phase-separated state; CDW: charge-density-wave state;
SP: singlet-pairing state.
Closed symbols and open ones show the phase-separated boundary and
the contour of $K_{\rho}=1$, respectively, for $N=10$ (upward triangles), 
$12$ (squares), $14$ (downward triangles) and $16$ (diamonds with a solid line).
The contours of other values of $K_{\rho}$ for $N=16$ are shown by solid lines.
The low-density limit of the phase-separated boundary~\cite{Ammon} 
is shown by a broken line.}
\label{fig.PD1D}
\end{figure}

We next calculate the spin gap
given by
$E_{\rm SG} = \lim_{N \to \infty} E_{\rm GS} (N; N_{\rm e}=N n)$,
where
$E_{\rm SG} (N; N_{\rm e}) = E_0 (N; N_{\rm e}; S_{\rm tot}^z = 1) 
- E_0 (N; N_{\rm e}; S_{\rm tot}^z = 0)$.
Here $S_{\rm tot}^z$ denotes the $z$-component of total spin for the system.
We estimate the magnitude of spin gap with use of
the data of $N=9,12,15$ for $n=2/3$,
$N=8,12,16$ for $n=0.5$, and $N=6,12,18$ for $n=1/3$.
The following fitting function is used for extrapolation to
$N \to \infty$~\cite{Ammon}:
\be
  \Gamma (N) = \Gamma (\infty) + A/N^m + B/N^{2m},\label{fittingfunction}
\ee
with $\Gamma = E_{\rm SG}$ and $m=1$.
Figure 2 shows the extrapolated values of the spin gap as a function of
$\alpha$.
At any electron density, the presence of spin gap
is clearly shown for the whole region of $\alpha > \alpha_{\rm c}$.
In the inset of Fig.\ 2, both horizontal and vertical axes are scaled by $\alpha$.
This figure makes the continuity to
a case of only ${\cal H}_{t3}$ (i.e., $1/\alpha = 0$)~\cite{Batista} clear.
The spin gap scaled by $\alpha$ has a peak near $\alpha=1$
at any filling.
\begin{figure}
\vspace*{0.5cm}
\epsfxsize=5.5cm
\centerline{\epsfbox{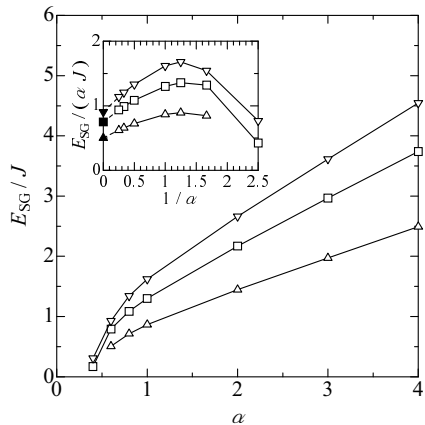}}
\caption{Spin gap versus $\alpha$ for the 1D $J$-$t_3$ model.
Each value is extrapolated from the finite-size data.
The values for $n=2/3, 1/2$ and $1/3$ are shown by
upward triangles, squares and downward triangles, respectively.
The inset shows the spin gap scaled by $\alpha$ as a function of $1/\alpha$.
The values of the analytic result for only the correlated hopping
($1/\alpha = 0$)~\cite{Batista}
are indicated by closed symbols.
}
\label{fig.spingap}
\end{figure}

An accurate way to estimate a spin-gap region is the singlet-triplet level
crossing method~\cite{AGSZ,ON}.
Nakamura et al. applied the method to the 1D {\it t-J} model,
assuming that the low-energy behavior of the model is described by
the effective Hamiltonian consisting the U(1) Gaussian model and
the SU(2) sine-Gordon model~\cite{NNK}.
One can check the validity by seeing that $x_r \equiv (x_{\rm ss} + 
3 x_{\rm st})/4$ is close to 1/2, where $x_{\rm ss}$ ($x_{\rm st}$) 
denotes the scaling dimension of the singlet (triplet) excitations.
These excitations are calculated under the boundary condition twisted from
the one for the ground state~\cite{NNK}.

When the level crossing method is applied to the $J$-$t_3$ model,
we should notice two points.
One is the presence of correlated hopping,
and the other is the absence of single-particle hopping.
Here we assume that 
in the low-energy region
the model given by
eq.\ (\ref{Hamiltonian}) is described by the above effective Hamiltonian
even for $t_3 > 0$.
The validity 
is checked from the scaling dimension to be shown later.
Meanwhile, the presence of single-particle hopping $t$
seems to be indispensable 
because it provides the finite gradient for linearization of the band dispersion.
To avoid the ambiguity, we begin with the finite value of $t$,
and then consider the limit of $t \to 0$.

Figure 3(a) shows the spin-gap boundary obtained by the level crossing method,
in the $\alpha$-$t/J$ plane at $n=0.5$ and $0.875$ in 16 sites.
There is no spin gap in the region above the solid line [case (A)],
while the remaining region should be the spin-gap region or the phase-separated
one [case (B)].
Obviously the limit of $t \to 0$ belongs to the case (B) at both electron densities.
The validity of the method is ensured in Fig.\ 3(b),
which shows that the scaling dimension $x_r$ in the gapless region ($t/J \gsim 0.8$)
is very close
to 1/2 ($0.502$-$0.504$)
for $N=16, n=0.5$ and $\alpha=1$.
One may ask whether 
the limit of $t \to 0$ and the case of $t = 0$ are connected
smoothly.
In Fig.\ 3(c) we show the magnitude of spin gap as a function of $t/J$.
The figure informs us no singularity of the spin gap at $t/J=0$.
This situation would hold also for other values of $\alpha$ and fillings
unless phase separation occurs.
Thus
we conclude that the $J$-$t_3$ model has a spin gap
in the whole region except the phase separated one~\cite{comment}.
\begin{figure}
\vspace*{1cm}
\epsfxsize=8cm
\centerline{\epsfbox{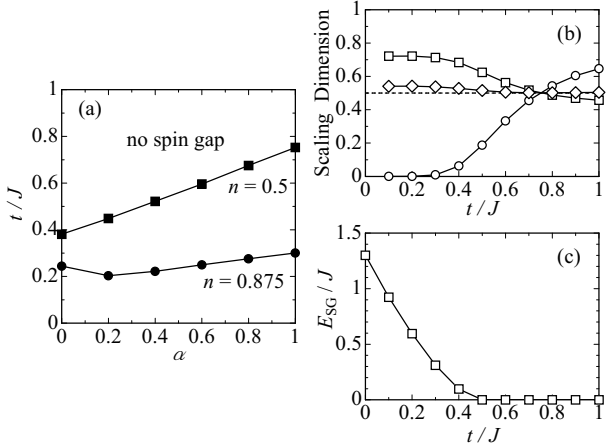}}
\caption{(a) Spin-gap boundary in the $\alpha$-$t/J$ plane at $n=0.5$ and $0.875$.
(b) Scaling dimension of the lowest singlet excitation ($x_{\rm ss}$) and
that of the triplet excitation ($x_{\rm st}$) for $N=16$, $n=0.5$ and $\alpha=1$.
The quantities $x_{\rm ss}, x_{\rm st}$ 
and $x_r \equiv (x_{\rm ss} + 3 x_{\rm st})/4$ are shown by
circles, squares and diamonds, respectively.
(c) Magnitude of spin gap as a function of $t/J$ for $n=0.5$ and $\alpha=1$.
Each value is extrapolated from the finite-size data ($N=8, 12, 16$).
}
\label{fig.spingapregion}
\end{figure}

Although the $J$-$t_3$ model has no single-particle hopping,
the charge velocity is not zero but finite 
due to 
the correlated hopping~\cite{Batista}.
Thus we expect that charge degrees of freedom for the system
behave as a Tomonaga-Luttinger (TL) liquid.
To support this,
we calculate the central charge $c$ which characterizes
the universality class of a model.
One can obtain $c$ by fitting the ground-state energy as
eq.\ (\ref{fittingfunction}) with $\Gamma = E_0 /N$ and $m=2$.
Here the coefficient $A$ is given by $- \pi v_{\rm c} c / 6$, and
the charge velocity $v_{\rm c}$ is extrapolated by
eq.\ (\ref{fittingfunction}) with $\Gamma = v_{\rm c}$ and $m=2$.
Note that $A$ does not include the spin velocity
due to the presence of spin gap.
With use of the data of $N=8, 12$ and $16$ with $n=0.5$,
we find that $c$ is almost unity ($c \sim 0.999$) in the region
$\alpha = 0.6$-$4.0$.
This indicates that charge degrees of freedom are described
by a TL liquid.
For $\alpha \lsim 0.4$, $c$ deviates largely from unity,
reflecting the phase separation which is no longer a TL liquid.

In the region $c = 1$,
one can use the following
universal formula about the correlation exponent
$K_{\rho}$:
\be
  K_{\rho} = \pi \sqrt{\frac{D n^2 \kappa}{2}},
\ee
where the Drude weight $D$ is defined as
\be
  D = \left. \frac{N}{2} \frac{\partial^2 E_0 (\phi)}{\partial \phi^2} 
  \right|_{\phi = 0}.
\ee
Here $E_0 (\phi)$ denotes the ground-state energy of the system under
the twisted boundary condition 
($c_{i+N,\sigma} = {\rm e}^{- {\rm i} \phi} c_{i \sigma}$).
When a spin gap is present, singlet-pairing
[charge-density-wave (CDW)] correlation functions
depend on the distance $r$ between sites as $r^{-1/K_{\rho}}$
[$r^{-K_{\rho}}$] for $r \gg 1$.
If $K_{\rho} > 1$,
singlet-pairing correlations
are the most dominant;
otherwise CDW correlations are the most dominant.
In Fig.\ 1 we show the contour lines of $K_{\rho}$ with $1.0$-$2.0$.

Our result for the values of $K_{\rho}$ is supported by direct calculation
of the equal-time pairing correlation function
$C(r) =
(1/N) \sum_i \langle P_i^{\dagger} P_{i+r} \rangle$.
Here the singlet pairing operator 
$P_i$ is given by
$P_i = (c_{i \uparrow} c_{i+1,\downarrow}
- c_{i \downarrow} c_{i+1,\uparrow}) / \sqrt{2}$,
and $\langle \cdots \rangle$ denotes the expectation value 
in the ground state~\cite{commentGS}.
Figures 4(a) and 4(b) show the distance dependence of 
$C(r)$ for various values of $\alpha$ with
$n=0.75$ fixed and that for various fillings with $\alpha=\infty$ fixed,
respectively.
In any case, 
pairing correlations reach up to long distance for
$K_{\rho} > 1$.
The sum $S = \sum_r C(r)$ provides the uniform (zero-momentum) component
of the pairing structure factor.
Figure 4(c) shows the result of $S$ as a function of $\alpha$
for various fillings.
The values of $\alpha$ at which $S$ has a peak are found to correspond to
the phase-separated boundary ($K_{\rho} = \infty$) in Fig.\ 1.
\begin{figure}
\vspace*{0.5cm}
\epsfxsize=7.5cm
\centerline{\epsfbox{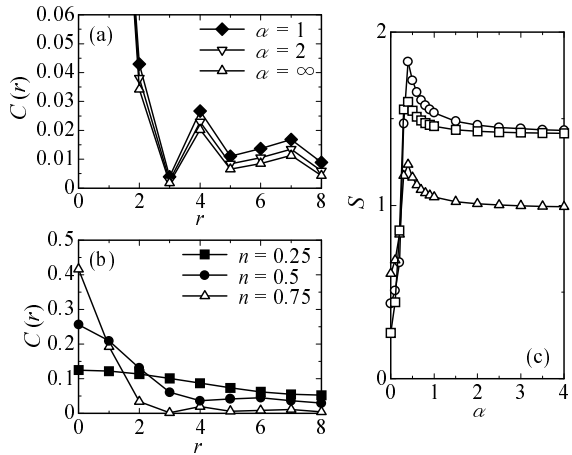}}
\caption{(a) Equal-time pairing correlation function as a function of
distance for the 1D $J$-$t_3$ model with $N=16$, $n=0.75$
and three different values of $\alpha$.
(b) Same as (a) but with $\alpha=\infty$ and three different values
of filling.
In (a) and (b), cases indicated by closed symbols are within
the singlet-pairing region ($1 < K_{\rho} < \infty$) in Fig.\ 1.
(c) 
Uniform component of the pairing structure factor 
as a function of $\alpha$ at $n=0.25$ (squares), 
$0.5$ (circles) and $0.75$ (upward triangles) in a 16-site cluster.
}
\label{fig.paircorr}
\end{figure}

The phase diagram consists of three regions: 
the phase-separated region, 
the CDW one and the singlet-pairing one.
Note that for $\alpha = \infty$ the CDW correlations are the most dominant
in the range $n > 2 - \sqrt{2} \simeq 0.586$~\cite{Batista}.
We emphasize that the pairing correlations are the most dominant
irrespective of $n$
for $\alpha=1$ corresponding to 
complete suppression of single-particle hopping 
in the strong coupling limit of the Hubbard model.

How is the situation changed on a square lattice?
We employ a 4$\times$4 cluster with the periodic boundary condition
in both $x$ and $y$ directions.
To search indications of superconductivity, we calculate equal-time
pairing correlations given by $C_{\rm s(d)} (\vec{r}) = (1/N) \sum_{\vec{i}} \langle 
\Delta_\pm^\dagger (\vec{i}) \Delta_\pm (\vec{i}+\vec{r}) \rangle$~\cite{commentGS,DR}.
The pairing operator $\Delta_\pm (\vec{i})$ is defined as 
$\Delta_\pm (\vec{i}) = c_{\vec{i} \uparrow} 
(c_{\vec{i} + \vec{\delta}_x,\downarrow} 
+ c_{\vec{i} - \vec{\delta}_x,\downarrow} 
\pm c_{\vec{i} + \vec{\delta}_y,\downarrow} 
\pm c_{\vec{i} - \vec{\delta}_y,\downarrow})$,
where $+$ ($-$) corresponds to extended-s-wave (d$_{x^2 - y^2}$-wave) symmetry,
and the unit vector in $x$ ($y$) direction is
represented by $\vec{\delta}_x$ ($\vec{\delta}_y$)~\cite{commentpair}.
In Fig.\ 5(a) we show the distance dependence of $C_{\rm s}(r)$ and $C_{\rm d}(r)$
with $r \equiv |\vec{r}|$ for $n=0.5$ and $\alpha=1$.
The extended-s-wave pairing correlations are stronger than the d$_{x^2 - y^2}$-wave
ones at all distances~\cite{YO}.
Figure 5(b) shows the extended-s-wave correlations for $n=0.5$ and various
values of $\alpha$.
The rapid decay at long distances for $\alpha=0.2$ reflects phase separation.
The behavior for $\alpha=1$ and $\alpha=\infty$ (i.e., only the correlated hopping)
is almost the same.
This means that the pairing correlations are hardly affected by the spin exchange
unless phase separation occurs.
Finally we show the uniform component of the pairing structure factor
$S_{\rm s(d)} = \sum_{\vec{r}} C_{\rm s(d)} (\vec{r})$ in Fig.\ 5(c).
For $n=0.25$ and $0.5$, we observe the abrupt increase in $S_{\rm s}$
as a function of $\alpha$.
This implies a transition from phase separation to s-wave superconductivity.
At fixed $\alpha (\gsim 0.4)$, as the electron density increases
the extended-s-wave correlations become weak.
Instead the d$_{x^2 - y^2}$-wave correlations become strong, 
but these do not excel the s-wave ones at least for $n \le 0.75$.
\begin{figure}
\vspace*{0.5cm}
\epsfxsize=7.5cm
\centerline{\epsfbox{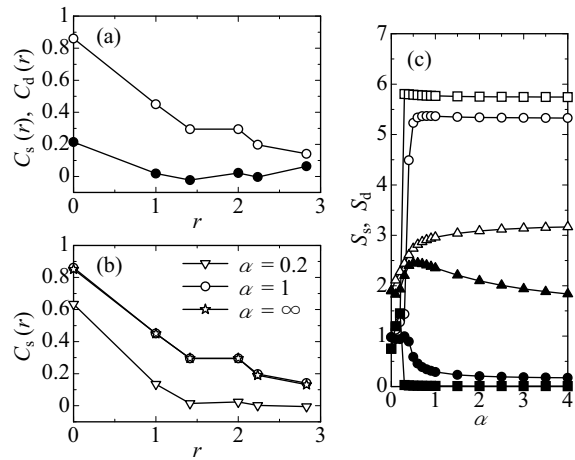}}
\caption{(a) Equal-time pairing correlation function as a function of
distance for the 2D $J$-$t_3$ model with $N=4\times4$, $n=0.5$
and $\alpha=1$.
Open (closed) symbols denote extended-s-wave (d$_{x^2 - y^2}$-wave) correlations
throughout Fig.\ 5.
(b) Extended-s-wave correlations with three different values of $\alpha$.
Other parameters are the same as in (a).
(c) 
Uniform component of the pairing structure factor 
as a function of $\alpha$ at $n=0.25$ (squares), 
$0.5$ (circles) and $0.75$ (upward triangles).
}
\label{fig.paircorr2D}
\end{figure}

In summary, we have found that the phase diagram of the 1D $J$-$t_3$ model
is composed of the three different regions.
Most of the parameter space is occupied by the singlet-pairing region.
Also in two dimensions, extended-s-wave superconducting states 
are likely to cover a wide range,
although we have not confirmed whether the long-range order exists.
A characteristic feature of the superconductivity discussed in this paper is 
to be caused by only the two-particle process.
This mechanism is essentially different from the Bardeen-Cooper-Schrieffer (BCS)
mechanism~\cite{BCS} which is based on the single-particle process and
an attractive potential between two electrons.

One of the authors (Y.S.) 
is grateful to M. Oshikawa, K. Okamoto and M. Kohno
for valuable discussions.
Y.S. is supported by JSPS Research Fellowships for Young Scientists.


\begin{thebibliography}{99}
\bibitem{comment} This statement cannot be confirmed in the limit $n \to 1$.
\bibitem{commentGS} The ground state is calculated in the subspace 
where the total momentum is zero.
\bibitem{commentpair} In this work we restrict hole (electron) pairs to nearest neighbors.
\end{thebibliography}
\end{document}